\documentclass[11pt,preprint,graphicx]{aastex}
\usepackage{graphicx}

\def\etal{\it et al. \rm }

\begin{document} 

\title{Stellar Populations and the Star Formation Histories of LSB Galaxies:
IV {\it Spitzer} Surface Photometry of LSB Galaxies}

\author{James M. Schombert}
\affil{Department of Physics, University of Oregon, Eugene, OR USA 97403;
jschombe@uoregon.edu}

\author{Stacy McGaugh}
\affil{Department of Astronomy, Case Western Reserve University, Cleveland, OH 44106;
stacy.mcgaugh@case.edu}

\begin{abstract}

\noindent Surface photometry at 3.6$\mu$m is presented for 61 low surface brightness
(LSB) galaxies ($\mu_o < 19$ 3.6$\mu$m mag arcsecs$^{-2}$).  The sample covers a
range of luminosity from $-$11 to $-$22 in $M_{3.6}$ and size from 1 to 25 kpc.  The
morphologies in the mid-IR are comparable to those in the optical with 3.6$\mu$m
imaging reaches similar surface brightness depth as ground-based optical imaging.  A
majority of the resulting surface brightness profiles are single exponential in shape
with very few displaying upward or downward breaks.  The mean $V-3.6$ color of LSB is
2.3 with a standard deviation of 0.5.  Color-magnitude and two color diagrams are
well matched to models of constant star formation, where the spread in color is due
to small changes in the star formation rate (SFR) over the last 0.5 Gyrs as also
suggested by the specific star formation rate measured by H$\alpha$.

\end{abstract}

\section{Introduction}

The primary characteristic of a galaxy is its luminosity and the deduced stellar mass
from that luminosity.  Of secondary interest is how that luminosity is distributed
(again, a proxy for the stellar mass distribution) and galaxy color, which reflect
the properties of the underlying stellar population.  The run of luminosity with
radius (surface brightness profiles) continue to be the most direct method of
deriving the size, luminosity and density scale parameters that are key to
understanding the details of galaxy formation.  The total stellar mass and its
distribution, as given by surface brightness profiles, also play important roles in
the various scaling relations between galaxy types.  For example, the Tully-Fisher
relation (TF; Tully \& Fisher 1977) is one of the strongest correlations in
extragalactic astronomy.  It provides a vital constraint on galaxy formation theory
(e.g., Eisenstein \& Loeb 1996; McGaugh \& de Blok 1998; Courteau \& Rix 1999; van
den Bosch 2000; Navarro \& Steinmetz 2000; Mo \& Mao 2000; 2004).

The dominant uncertainty in the TF relation is the deduction of stellar mass from
luminosity.  In order to constrain the luminosity to stellar mass relationship, our
stellar population models agree that longer wavelength observations more accurately
map into stellar mass (Bruzual \& Charlot 2007).  Empirically, the scatter in the
Tully-Fisher relation declines as one goes from blue to red to NIR wavelengths
(Verheijen 2001), consistent with the expected decrease in scatter in $M/L_*$.
However, LSB galaxies are, by definition, very faint in luminosity density, typically
well below the natural sky brightness.  This is particularly a problem at near-IR
wavelengths where the sky brightness from the ground are several magnitudes brighter
than the sky at near-UV wavelengths.  To this end, the longer wavelength IRAC data
from the {\it Spitzer} orbiting telescope provides a unique probe of stellar mass,
dust and star formation, useful for testing whether extinction or fluctuations in the
star formation rate cause deviations from the TF relation.

The goal of this paper, the fourth in our series to understand the star formation
history of LSB galaxies, is to present the 3.6$\mu$m imaging for a sample of HSB and
LSB galaxies obtained during the 2009/2010 observing seasons.  The data was acquired
as part of a study of the baryonic Tully-Fisher relation (McGaugh \& Schombert 2013)
where the total luminosity of a galaxy at 3.6$\mu$m provides a more accurate measure
of stellar mass and an independent estimate of the color-mass to light ratio.  During
data reduction it was determined that the images were comparable in depth to optical
imaging, and open to a full surface photometric analysis for direct comparison to
optical values obtained in Paper I (Schombert, Maciel \& McGaugh 2011).

\section{Observations}

\subsection{Sample Properties}

Selection criteria based on magnitudes preferentially miss LSB galaxies because much
of their integrated light resides below the detection threshold of wide area surveys
(McGaugh 1994).  However, the low sky afforded by {\it Spitzer} observations is
ideally suited to the observation of LSB galaxies and are vastly superior to ground
based observations in the $JHK$ bands, which still lag well behind other types of
data for these systems because of the obvious technical challenges.  To this end, our
sample was designed to explore a wide swath of under-sampled parameter space in the
mid-IR, probing a large region in galaxy mass and gas fraction as well as surface
brightness.

The sample presented herein are a combination of LSB galaxies, selected for {\it
Spitzer} cycle 9 observing based on their central surface brightness and existing
optical and H$\alpha$ imaging, and a small subset of comparison HSB galaxies.
The HSB galaxies were taken from McGaugh (2005), a selected for their large mass and
observed for a baryonic Tully-Fisher cycle 5 program.  All the LSB galaxies are
selected from the Schombert F and D LSB catalogs (Schombert \& Bothun 1988; Schombert
\etal 1992; Schombert \etal 1995) with some additional, previously known, LSB UGC
galaxies.  All the LSB galaxies have central $B$ surface brightness $\mu_o > 23$ mag
arcsecs$^{-2}$, which differs from previous {\it Spitzer} LSB programs which usually
observed intermediate surface brightness galaxies ($22 < \mu_o < 23$ mag
arcsecs$^{-2}$).  The basic characteristics of the entire sample are found in Tables
1 and 2.  All the photometry (found in Tables 3 and 4) presented herein were
corrected for Galactic extinction using the extinction maps from Schlegel \etal
(1998) and the extinction curve of Li \& Draine (2001).  Redshifts are determined
from previous HI work (Eder \& Schombert 2000; Schombert \etal 1992) and used to
derive distances based on the CMB reference frame or tabulated in the Extragalactic
Distance Database (Tully \etal 2009).  Stellar and gas masses, plus other optical
values (such as $L_{H\alpha}$), are taken from Paper I (Schombert, Maciel \& McGaugh
2011).

\begin{figure}[!ht]
\centering
\includegraphics[scale=0.80,angle=0]{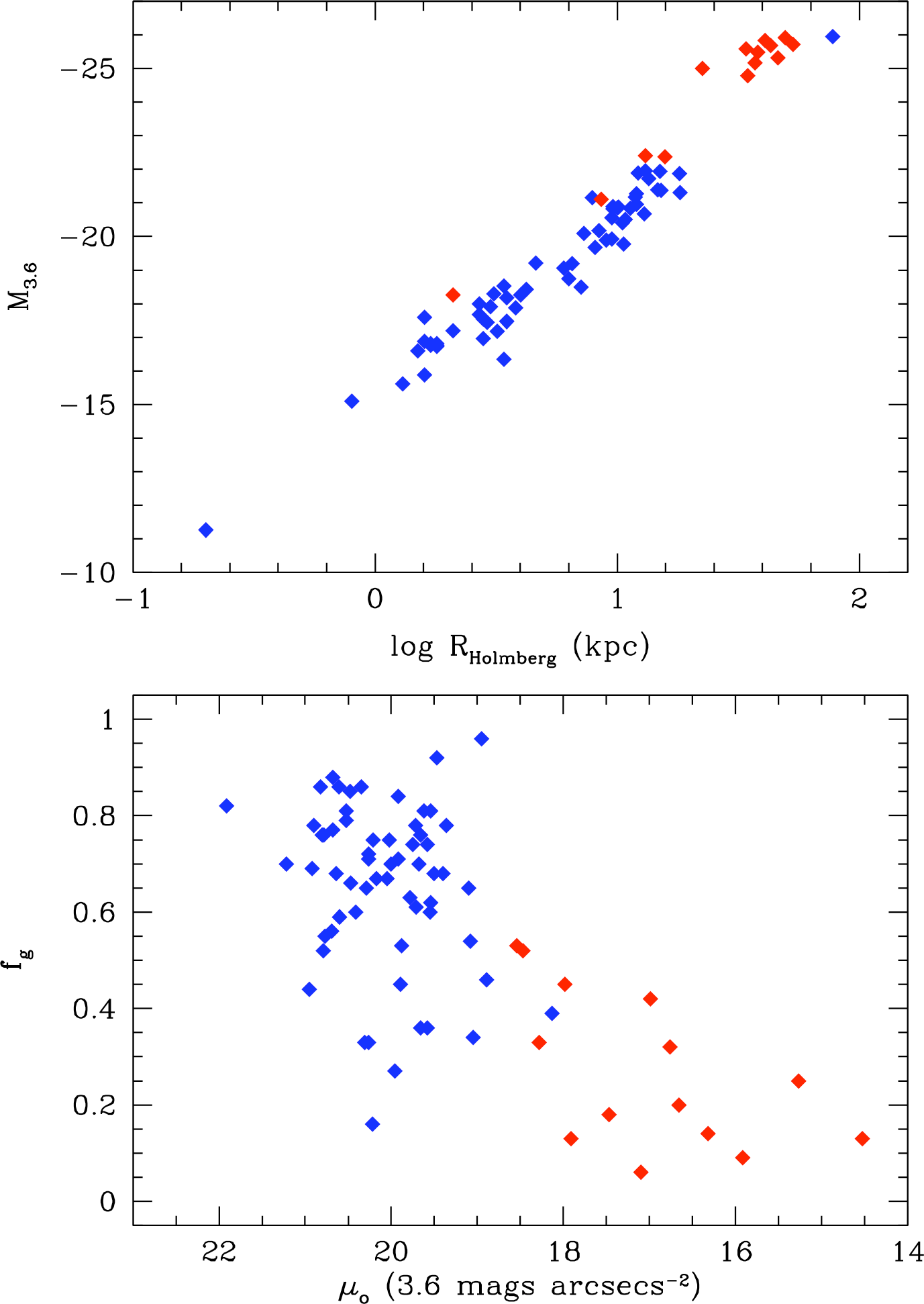}
\caption{\small The top panel displays absolute 3.6$\mu$m magnitude versus the
3.6$\mu$m Holmberg radius.  The blue symbols are the LSB galaxies and the red symbols
are the HSB sample from our baryonic Tully-Fisher project (McGaugh \etal
2010).  The largest galaxy in the sample is F568-6, a supergiant Malin cousin, the
smallest is Cam B.  The linear relationship between size and luminosity is evident.
The bottom panel display central surface brightness ($\mu_O$, based on exponential
fits to the surface brightness profile) versus gas fraction
($f_g = M_{gas}/M_{gas}+M_*$).  While presenting a wide range in $f_g$, the highest
$f_g$ galaxies are low in surface brightness.  Galaxies with $f_g > 0.5$ are defined
as "gas-rich".
}
\label{mag_rad_fg}
\end{figure}

The range in absolute 3.6$\mu$m magnitude and galaxy size can be found in the top
panel of Figure \ref{mag_rad_fg}.  Determination of these parameters is described in
\S3.1, and comparison to optical values can be made with the knowledge that the mean
$V$-3.6 color is 2.3 for disk galaxies.  The LSB galaxies range from $-$10 to $-$22
$M_{3.6}$, whereas the HSB galaxies range from $-$18 to $-$26.  As the HSB subset was
selected for their high rotation values (i.e. high mass) to explore the baryonic
Tully-Fisher relation, it is not surprising to find that they fill the bright end of
the sequence.   The brightest galaxy in our sample is F568-6, a supergiant Malin
cousin (Bothun \etal 1990).  Size, as given by the Holmberg radius, scales with total
magnitude, such that the smallest galaxy in the sample (Cam B) has a radius of 0.2
kpc.  The typical LSB galaxy ranges from 1 to 10 kpc, which encompasses dwarfs and
normal sized disks.  Most of the HSB galaxies are greater than 30 kpc, i.e., Milky
Way sized systems, again due to their selection for high mass.

The bottom panel in Figure \ref{mag_rad_fg} displays one of the primary differences
between LSB and HSB, the gas fraction defined as the gas mass of a galaxy (HI plus
H$_2$ and He) divided by the total mass of the galaxy in stars and gas (i.e.
baryons).  Figure \ref{mag_rad_fg} displays what is well known about LSB galaxies
(McGaugh \& de Blok 1997, Schombert \etal 1995) that $f_g$ rises with lower central
surface brightness, although there is only a weak correlation between $f_g$ and
$\mu_o$ for there exist several LSB galaxies with gas fractions similar to HSB
galaxies.  We define the phrase `gas-rich' for those galaxies with $f_g$ values above
0.5 and a majority of LSB galaxies (81\%) are gas-rich.  Thus, while LSB galaxies are
difficult to detect at wavelengths were stellar light dominates, they are often quite
bright at 21-cm, i.e. neutral hydrogen (Schombert \etal 1992).

\subsection{Detector Characteristics}

The images used in this study were obtained with the {\it Spitzer} InfraRed Array
Camera (IRAC).  Briefly, IRAC is a four-channel camera that provides simultaneous 5.2
by 5.2 arcmin images at 3.6, 4.5, 5.8 and 8 $\mu$m. Two adjacent fields of view are
imaged in pairs (3.6 and 5.8$\mu$m; 4.5 and 8.0 $\mu$m) using dichroic beamsplitters.
All four detector arrays in the camera are 256 by 256 pixels in size, with a pixel
size of 1.2 by 1.2 arcsecs.

Our primary data was taken in the 3.6$\mu$m band, whose filter centers at 3.55$\mu$m
and covers between 3.1 and 3.9$\mu$m a nearly constant transmission.  The maximum
exposure time of 100 secs was used for each observation, with 24 frames taken for
each galaxy for a total of 2,400 secs of integration per pixel.  The frames were
obtained using a 12-point Reuleaux pattern in a 1/2 subpixel dither.  The resulting
pixels were 0.61 arcsecs in resolution.  The frames were flatfielded and calibrated
using the standard {\it Spitzer} pipeline.  As all our objects were small relative to
the field of view, no geometric or spatial flux corrections were applied.  The FWHM
of the PSF was 1.7 arcsecs for 3.6 and 4.5$\mu$m detectors.  No corrections were made
for the pixel phase effect as our galaxies are much larger in angular size than any
inter-pixel effects.   The diffuse light component is removed by our sky procedures.

Photometric calibration is supplied by the {\it Spitzer} processing pipeline and has
an estimated zeropoint error of less than 2\% (Reach \etal 2005).  However, the true
photometric error will be strongly dependent on the knowledge of the correct sky
values for each frame, as has always been true for low surface brightness photometry
(Schombert, Maciel \& McGaugh 2011).  The sky brightness error will have two major
components, the pixel-to-pixel noise in the detector pixels and the flatness of the
image on the scale of the object to be studied.

As has been successful in our past surface photometry work (Schombert 2011), we have
used a sky box technique to determine the absolute sky value and its uncertainty.
This technique uses visually selected regions of the galaxy image that are free of
contaminating stars and background galaxies.  Typically 10 to 20 boxes of 20x20
pixels are used in these calculations.  For 90\% of the galaxy frames in our sample,
the sky brightness varied between 20 and 23 AB mags with a mean of 21.0 (where a 0 AB
magnitude object has a flux of 280.9 Jys at 3.6$\mu$m).  The mean error on the sky
was 2.2\%, but 80\% of the sample had a sky noise of less than 3\%.  This error
dominates all other sources of error in our surface photometry and aperture
magnitudes, and is used to assign the uncertainty in all photometric values.

The mean sky value of 21 mag arcsecs$^{-2}$ at 3.6$\mu$m is 2.5 mags darker than the
optical night sky at our best ground observatories.  Thus, where the best optical
surface photometry reaches a depth of 28 to 29 $V$ mag arcsecs$^{-2}$ (Schombert
1988), it is not uncommon in this sample for the mid-IR data to reach 25 mag
arcsecs$^{-2}$ at 3.6$\mu$m with errorsbars of less than 0.5 mags (this corresponds
to 0.4 $L_{\sun}$ pc$^{-2}$).  This is equivalent to optical photometry, but at least
4 mags fainter than ground-based IR surface photometry (e.g., Galaz \etal 2002).

\subsection{Frame Cleaning}

\begin{figure}[!ht]
\centering
\includegraphics[scale=0.80,angle=0]{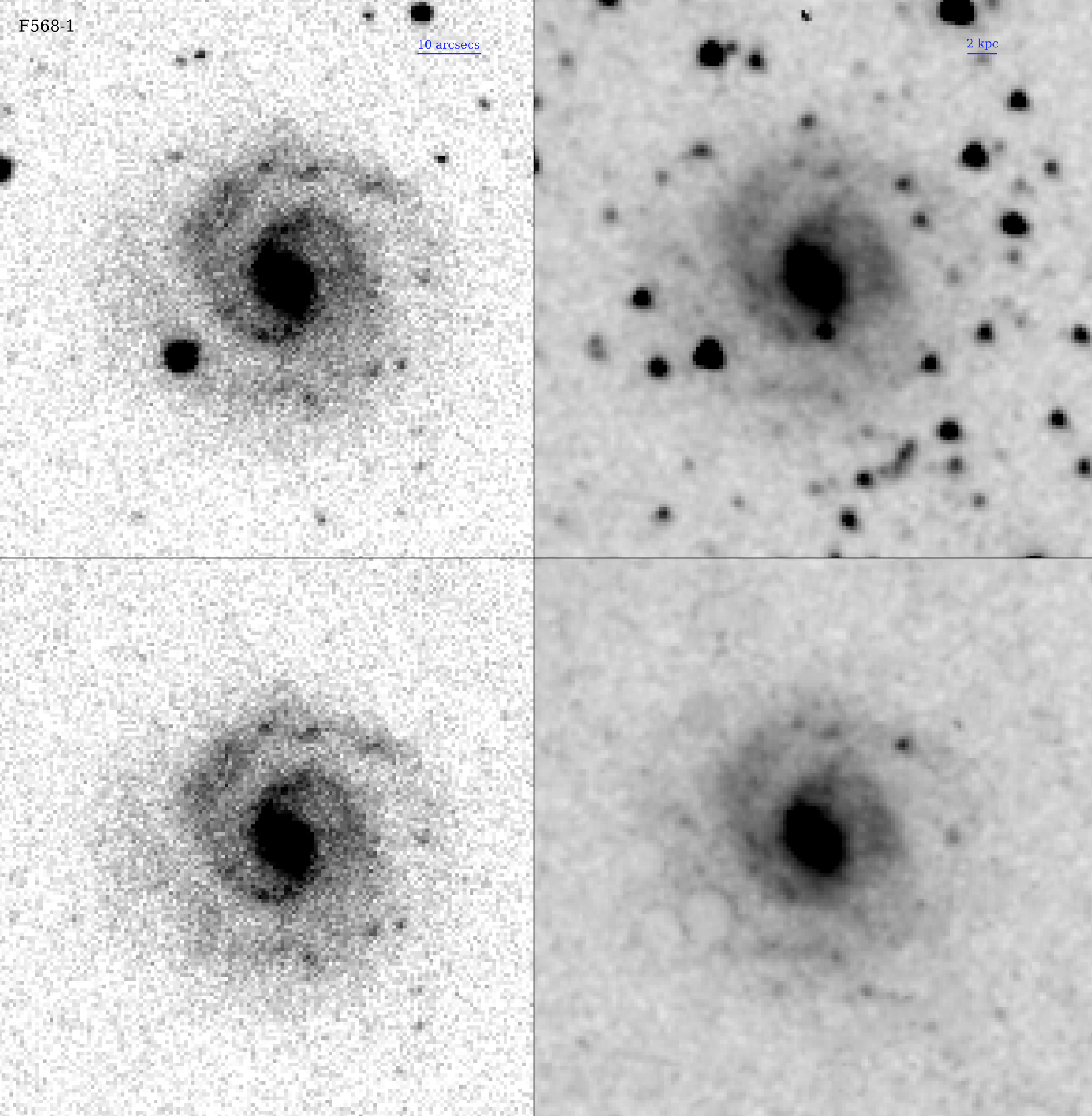}
\caption{\small Optical $V$ (left) and {\it Spitzer} 3.6$\mu$m (right) images for LSB
galaxy F568-1.  The $V$ image is a 600 sec exposure from KPNO's 2.1m (plate scale of
0.6 arcsecs per pixel), the 3.6$\mu$m image is based on a 2400 secs exposure of the
same plate scale.  The top panels display the uncleaned images, note the larger number
of point sources in the 3.6$\mu$m frame (background galaxies) compared to the $V$
frame.  The bottom panels display the results of the point source cleaning algorithm which
are design to clean objects not associated with structure in the galaxy itself.  All
features associated with H$\alpha$ emission were preserved.
}
\label{nostars}
\end{figure}

Perhaps the most salient difference between optical and {\it Spitzer} imaging is the
sharp increase in the number of point sources, not associated with the galaxy of
interest, visible in each frame. Figure \ref{nostars} displays a comparison of
between 150 sec $V$ image (taken with the KPNO 2.1m) and a 2400 sec {\it Spitzer}
IRAC 3.6$\mu$m image.  The number of point sources in the 3.6$\mu$m image is a factor
of 10 greater than the number in the $V$ image, although this is expected from early
{\it Spitzer} number counts (Fazio \etal 2004).  We have matching HST WFC3 imaging of
one object in our sample (F415-3), and comparison with those images reveals that 1)
bright sources in common with the $V$ and {\it Spitzer} images are mostly galactic
stars or unresolved nearby galaxies, and 2) faint sources found only the 3.6$\mu$m
images are background galaxies, either resolved in the WFC3 frames or sufficiently
faint as to be below any galactic star magnitude limit.

While the percentage of the image frame contaminated by point sources is still a
small fraction of the total number of pixels (typically less than 10\%), there is a
much higher probability that a significant portion of the galaxy image has
contaminating point sources compared to $V$ images (although it begs the question
that these same sources are interfering with $V$ images in a fashion that is not
visible in the $V$ frames).

There are three salient issues for the larger number of contaminating point sources
in the {\it Spitzer} images; 1) how the point sources interfere with isophote
fitting, 2) how much of the galaxy luminosity is contaminated by stellar objects and
3) how does the observer distinguish between unresolved galaxy features (clusters,
HII regions, etc.) and contaminating point sources.  The first issue can be resolved
by masking the more obvious stellar features, then allowing the ellipse fitting
algorithms to automatically remove pixels above and below a set threshold.  Even in
small LSB galaxies, there are a sufficient number of remaining pixels to determine
the mean isophote value.

The second issue can be mitigated by using the measured isophote values to re-fill
the masked pixels with galaxy light.  For small masked areas in the outer regions,
this is a simple process with little increase in the uncertainties on the aperture
magnitudes.  Large masked areas in the core region are most problematic.
Fortunately, most LSB galaxies are more symmetric in the core regions, such that
filled in masked areas appear visually to be sufficient.  Galaxies with large
contaminating objects in their cores were simply discarded from our sample.

The third issue, discriminating foreground stars (or background galaxies) from real
point-like objects in the galaxies (e.g., unresolved clusters or knots), is the most
difficult to replicate in an automatic script.  For this study we have followed three
guidelines; 1) any feature associated with an enhanced region of surface brightness
(i.e., a spiral arm or bulge) was not removed, 2) any feature visible in our deep $V$
or H$\alpha$ frames was not removed and 3) any feature which displayed a `soft' core
(suggesting a non-stellar profile) was not removed.  Operationally, there is no
simple method to automatically apply the above criteria, so we allowed the processing
pipeline remove objects beyond the 5\% isophote, but used visual inspection for the
inner regions.

The results are fairly successful as can be seen in Figure \ref{nostars}.  The bottom
images are the cleaned and filled frames for $V$ and {\it Spitzer} 3.6$\mu$m.  There
is very little obvious evidence of the removed areas, and there is no disjoint
increase or decrease in the aperture magnitudes curve-of-growth that would signal an
error in the filled regions.  However, at 3.6$\mu$m, these corrections involve a
significant fraction of the galaxy light, up to 15\%, and makes comparison with other
studies difficult.  This technique is much more conservative than the methods applied
by Munoz-Mateos \etal (2009).  For a comparison of their cleaned images (their Figure
2) indicates a number of faint point sources that remain after cleaning.  As their
SINGS sample studies high surface brightness spirals and irregulars, their slightly lower
luminosities should not significantly contribute to their error budget, but may be
critical in our LSB sample.

\subsection{Isophotometry}

Determining mean isophotes followed the same procedures that we have applied to
optical data for LSB galaxies (Schombert, Maciel \& McGaugh 2011).  Frames that were
manually cleaned were submitted to the ARCHANGEL pipeline (Schombert 2007) in the
same manner as optical data.  Sky boxes were used to determine the local sky in each
frame.  Ellipse fitting was performed using the standard Fourier series iterative
least-squares algorithm.  Given the irregular morphology of LSB galaxies, very few of
the isophote contours are truly elliptical.  However, to first order, a round shape
with a long and short axis is the best approximation and the errors in the surface
photometry are dominated by knowledge of the sky value rather than RMS around each
ellipse.  All the data (images, surface brightness profiles, colors and fit
parameters are found at our website, http://abyss.uoregon.edu/$\sim$js/spitzer).

\begin{figure}[!ht]
\centering
\includegraphics[scale=0.60,angle=0]{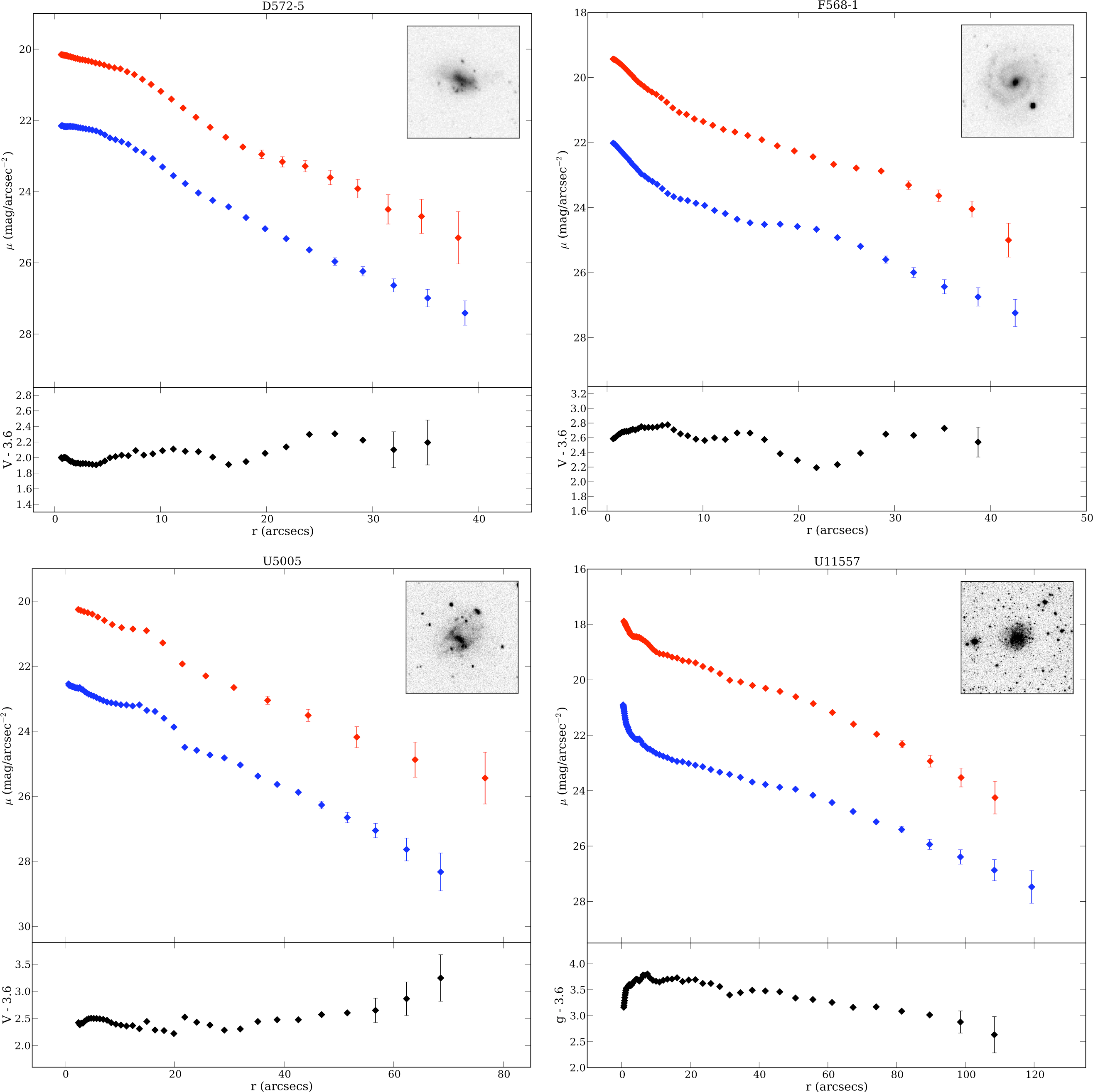}
\caption{\small A comparison of four surface brightness profiles in $V$ (blue) and
3.6$\mu$m (red).  The resulting $V-3.6$ color profiles are also shown.  The depth of
the 3.6$\mu$m data is compatible to the optical data, and all the features in the
optical profiles are reproduced in the 3.6$\mu$m profiles despite the broad range in
morphology.
}
\label{sfb_profiles}
\end{figure}

A few selected surface brightness profiles are shown in Figure \ref{sfb_profiles} to
display the range in the sample in terms of size and central surface brightness.  The
surface brightness profiles for the entire sample are available for download from our
website.  While some LSB galaxies have a clear bulge and disk appearance (plus a
double-horned HI profile), which signals a rotational dominate oblate 3D shape, many
LSB galaxies are irregular in appearance (some with single-horned HI profiles) with
no indication that they are oblate, prolate or triaxial in 3D shape (Sung \etal
1998).  However, historically, LSB surface brightness profiles are displayed as the
major axis ($r$) versus surface brightness ($\mu$, mag arcsecs$^{-2}$) and fit with
exponential fitting functions assuming a flattened shape.  We simply follow past
procedure, but make the reader aware that this does not assume a disk-like structure
for all LSB galaxies (Schombert \etal 1997).

Aperture and total magnitudes are determined from the same procedures as the optical
data for LSB galaxies (Schombert, McGaugh \& Maciel 2013).  Aperture magnitudes used
the best fit ellipses for the radius in question.  Luminosities were summed for all
the unmasked pixels interior to the ellipse (a surveyor method was used to include
the partial pixels at the ellipse edge).  Masked pixels were replaced with the mean
surface brightness of the fitted isophote for that radius.  The amount of galaxy
light derived from masked pixels varied from 2 to 15\%.  Total magnitudes were
derived from asymptotic fits to the curve-of-growth.  Beyond the 90\% total
luminosity radius, the apertures were replaced by the mean isophote value in order to
capture the light associated with the very LSB halos.

This adjusted curve-of-growth procedure is highly successful for the LSB galaxies in
the sample as all the galaxies have well-defined curves-of-growth at the faint ends
that convert to a clear total magnitudes.  The errors quoted in Tables 3 and 4 reflect the
Poisson noise for the galaxies, error in the sky value and errors in the asymptotic
fit.  Of the three sources of error, Poisson noise dominates the total magnitude
calculations by a factor of ten.  While the errors are atypically low, comparison
with other studies will reveal differences due to luminosity determination techniques
rather than an external check to the luminosities.  For example, comparison to SING
luminosities (Munoz-Mateos \etal 2011), reveals an external scatter of 5 to 10\%.

Galaxy size is determined by the isophotal Holmberg radius (Holmberg 1958; Faber \&
Gallagher 1979).  This is the size of the major axis at the point where the surface
brightness profile reaches 26.5 $B$ mag arcsecs$^{-2}$.  Assuming a $B-V$ color of
0.5 and a $V$-3.6 color of 2.5 produces a surface brightness cutoff of 23.5 mag
arcsecs$^{-2}$ at 3.6$\mu$m.  However, using this isophote only captures 50 to 60\%
of an LSB galaxies light (due to color gradients).  Therefore, we reduced the
isophote value to 24.5, which captures typically 80\% of a galaxy's light.  This
isophote is used to calculate the Holmberg radii ($R_{H}$) quoted in Tables 3 and 4,
the major axis of the galaxy at the 24.5 3.6$\mu$m isophote.  As displayed in Figure
\ref{mag_rad_fg}, the galaxy sizes in the sample range from dwarfs (1 to 2 kpcs) to
large disks (10 to 20 kpcs).  The Holmberg radius is strongly correlated with disk
scalelength ($\alpha$, based on exponential fits to the profile) such that
2.5$\alpha$ is equivalent to a galaxy's Holmberg radius.

Comparison to optical colors used either Johnson $V$ frames obtained from the KPNO
2.1m (Schombert, Maciel \& McGaugh 2011) or from SDSS $g$ frames taken from DR7
(Abazajian \etal 2009).  The SDSS $g$ values were converted to Johnson $V$ using the
standard SDSS conversions (Jester \etal 2005).  Colors ($V/g$ - 3.6$\mu$m) were
calculated by three methods; 1) direct subtraction of the aperture magnitudes
(integrated colors), 2) differences in the surface brightness $V$ and 3.6$\mu$m
profiles and 3) differential magnitudes (annular apertures).  Of the three methods,
the best total colors were provided by direct comparison of the asymptotic
magnitudes.  The best measure of spatial color was the differential surface
brightness profiles, examples of which are shown in Figure \ref{sfb_profiles}.  Color
gradients were determined using interpolation of the surface brightness profiles.
Again, the errors were dominated by sky value in both $V$ and 3.6$\mu$m and were
added in quadrature for the errors on the quoted colors.

\section{Discussion}

\subsection{Optical to Mid-IR Surface Brightness Profiles}

The most important result from our study is that the {\it Spitzer} observations reach
to similar depth and radii as the deepest optical surface brightness profiles, in
many cases farther than previous optical work of similar exposure times.  This type
of accuracy in the IR is simply not obtainable from the ground due to the high
atmospheric absorption combined with a bright background.  Therefore, IR space
imaging has numerous advantages for studies of galaxy mass since a mid-IR luminosity
minimizes the internal absorption corrections, resulting in photometric parameters
that better reflect the underlying stellar population (McGaugh \& Schombert 2013).
In addition, the morphology and structure of a galaxy in the mid-IR will tend to
follow the underlying kinematic stellar distribution, rather than being distorted by
dust lanes and recent star formation.

The visual morphology of the LSB galaxies fall, primarily, into the extreme late-type
classes (Sm, Im or dI).  Only two of the dwarf LSB galaxies (D-class) have any
symmetric shape (Sm class) and only seven of the F-class LSB galaxies have Sc or Sd
morphology.  The appearance of all the LSB galaxies at 3.6$\mu$m is identical to
their optical appearance (see Figure \ref{nostars}).  This was not unexpected as
optical color maps reveal very little absorption by dust or gas, so extinction that
changes the morphology of early-type spirals from optical to IR has little effect on
LSB galaxies.  This also reinforces expectations that, with low star formation rates
(SFR's), LSB galaxies will vary little in stellar population age with spatial
position, unlike spirals.

Surface brightness profiles are found in Figure \ref{sfb_profiles}, and visual
inspection of the side-by-side surface brightness profiles reveals that the optical
and mid-IR profiles follow the same slopes and contain the same general features
(bulges, lens and disk length).  The few differences in the optical and mid-IR
profiles can be attributed to asymmetric features (i.e., star-forming regions, see
\S3.1) which can have different luminosities at the optical and IR wavelengths
depending on the age of the stellar population.  This result was also expected based
on the behavior of multi-wavelength profiles from the SINGS project (Munoz-Mateos
\etal 2009, see their Figure 6) where strong differences in profile shape were noted
between the near-UV and far-IR, but with little change from optical to the mid-IR.

It is standard procedure to fit late-type galaxies to exponential profiles, for one
the most distinguishing characteristics between ellipticals and spirals is a
power-law versus exponential surface brightness profile (Schombert 2013).  The
consistency of these shapes as a function of morphology suggests that this is a
property that is imposed during galaxy formation and tied to physical properties,
such as total angular momentum.  With respect to disk galaxies, exponential profiles
in the outer regions are well defined by two parameters, central surface brightness
($\mu_o$) and scalelength ($\alpha$, van der Kruit 2002).  Central surface brightness
is only loosely correlated with global galaxies properties, such as galaxy mass.
However, scalelength, which is independent of Hubble type, increases with galaxy mass
(de Jong 1996; Fathi 2010).

The fitting an exponential profile to an irregular galaxy often has subtle
differences compared to procedures for spirals.  Very few of the LSB galaxies in our
sample have a well defined bulge+disk appearance (e.g., F568-1, in Figure
\ref{nostars}, F579-1 and UGC 11557).  Of the 59 galaxies in the sample classified as
LSB, only 15 have canonical bulge+disk morphology, although this is not a statement
on the morphology of LSB galaxies as a type of galaxy as all the D class galaxies in
our sample were selected for irregular morphology and specifically avoiding a
disk-like appearance.  Without a well defined bulge, the location of the isophotal
center becomes an exercise of the region of highest surface brightness or the
geometric mean from out isophotes.

While symmetric disk galaxies are not dominant in our sample, even an irregular LSB
galaxy tends to have a brighter central region surrounded by a fainter envelope.
This central region is rarely at the same center as the outer isophotes, but is
typically within a half of a scalelength ($\alpha$) of the mean isophotal center.
Truly undefined objects (e.g., D500-4, F565-V2, UGC 5209) are rare, as are galaxies
with nearly constant interior surface brightness than a sharp drop-off (box profile,
e.g., D500-3, D572-5, ESO215-G009).

Despite their irregularities, most of the surface brightness profiles at 3.6$\mu$m
are adequately described by an exponential fit.  We divided the sample by profile
shape in three categories: 1) box-like shape (flat core region with an exponential
drop-off), 2) disk (pure exponential) and 3) bulge+disk (two distinct components, the
bulge need not be a r$^{1/4}$ power-law shape as is common with bright spiral
bulges).  These classifications are listed in Tables 1 and 2.  Half the LSB sample have
profiles which described by a single exponential (disk-like), which due to the close
correspondence to the optical profiles, matches previous results (Schombert, McGaugh
\& Maciel 2013).  The other half of the sample was evenly divided into box-like and
bulge+disk shapes.  Unsurprisingly, the box-like profiles are associated with most
irregular morphologies and the least elliptical isophotal shape.

We have also classified each profile according to the classification scheme proposed
by Erwin \etal (2008), where a Type I profile has has no breaks from an exponential,
Type II has a downward break, and Type III has an upbending break.  Only nine (15\%)
galaxies were classed as Type II or III, which is significantly different from
Herrmann, Hunter \& Elmegreen (2013) who found 77\% of their dwarfs to have Type II
or III profiles.  For our dwarf galaxies ($R_{25} < 10$ kpc) our Type II or III
numbers increase to 30\% of the sample.  Part of the difference is due to stylistic
differences in applying the classification scheme.  For example, a box-like profile
would automatically be a Type II profile, however, if the flattened profile is clearly
an interior phenomenon and unrelated to the exponential fit in the halo, we called
this a Type I.  Part of the difference is also due to the low surface brightness
nature of our sample, as the outer isophotes have less S/N than higher surface
brightness systems.  Larger errorbars would disguise any break.

\begin{figure}[!ht]
\centering
\includegraphics[scale=0.80,angle=0]{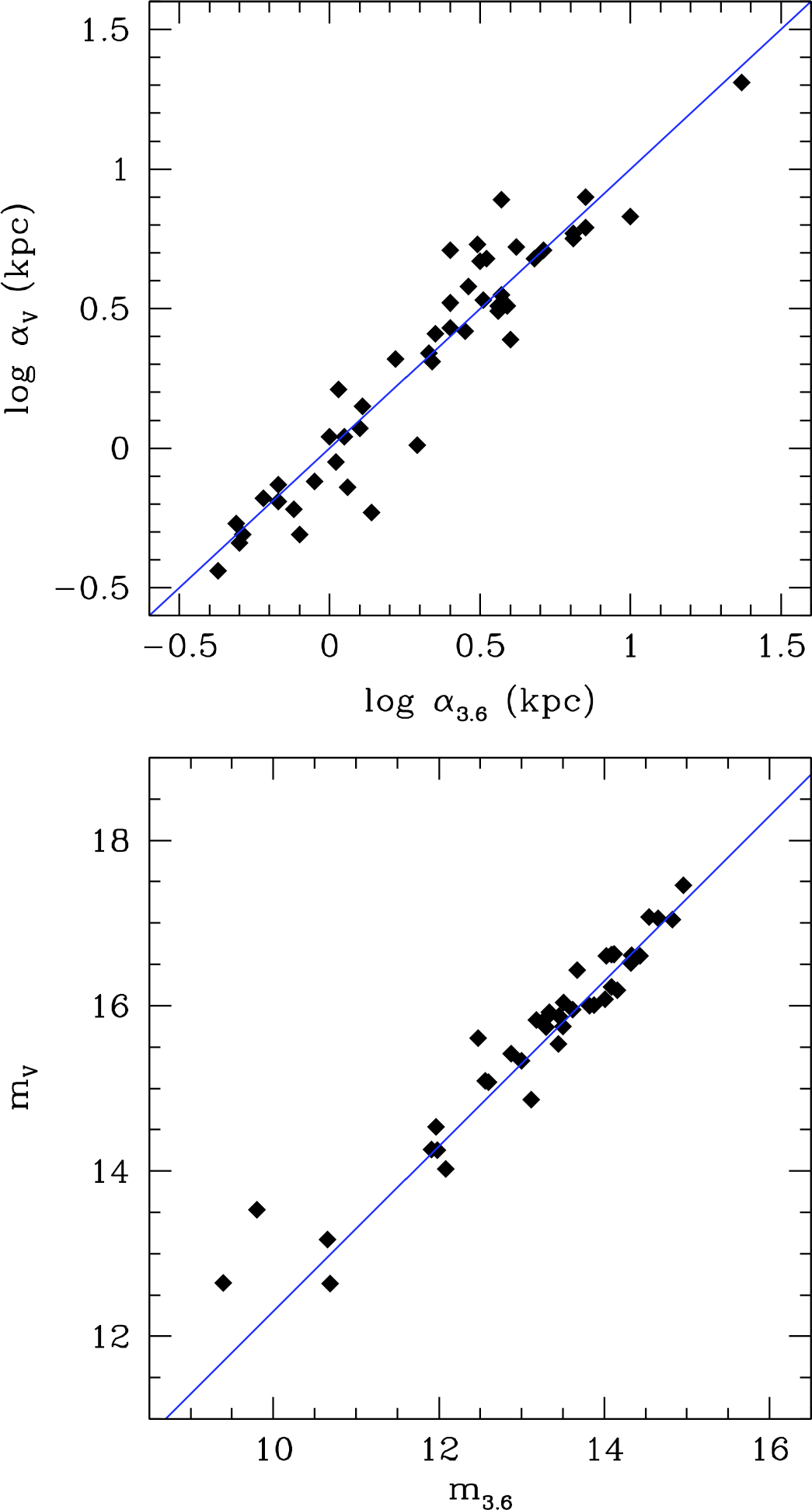}
\caption{\small Comparison of size and luminosity between optical and mid-IR imaging.
The top panel display a comparison of scalelength, $\alpha$, based on exponential
fits where the blue line is the unity relationship.  Given the similarity between the
optical and mid-IR surface brightness profiles, the close correspondence between
scalelength fits is unsurprising and reinforces the universality of the exponential
shape for late-type galaxies.  The comparison of total apparent magnitude is shown in
the bottom panel, the unity line assumes a $V-3.6$ color of 2.3.  The correspondence
is excellent considering the range in color for LSB galaxies.
}
\label{scalelengths}
\end{figure}

Figure \ref{scalelengths} displays a comparison of scalelength, $\alpha$, obtained by
exponential fits to the 48 galaxies in our sample with both optical ($V$ or SDSS $g$)
and mid-IR (3.6$\mu$m) profiles.  The correspondence is excellent considering the
differences in wavelength, telescopes, detectors and sky background.  The total
radial extent of the stellar component of a LSB galaxy is well described either by
optical or near-IR imaging.  However, space mid-IR imaging displays greater S/N due
to the fact that the luminosity of older stars peaks in the mid-IR and the sky is
darker than the ground.  There is a slight tendency for large galaxies to be
under-sized at 3.6$\mu$m and small galaxies to be over-sized (compared to their $V$
scalelengths), but the trend is not statistically strong.

We conclude that the exponential profile shape for LSB galaxies is universal at
3.6$\mu$m, even given the problematics due to irregular isophotes.  There is no
compelling reason to conclude any other fitting function is a better fit,
particularly considering the uniformity in profiles from optical $V$ to the mid-IR.
Given the range in kinematics for the sample, from rotating disk with flat rotation
curves to triaxial irregulars with solid-body motion, the mechanism determining
galaxy structure must be ubiquitous.  Also, given the similarity in shape between HSB
and LSB galaxies, star formation must not play a dominant role (Ferguson \& Clarke
2001).  Simulations of gravitationally self-consistent disk collapse naturally
produce exponential shapes (Mestel 1963; Dalcanton, Spergel \& Summers 1997),
however, the exact mechanism is unclear.  Features such as bars and spiral arms serve
to redistribute angular momentum (Hohl 1971) and tend to form double exponentials as
seen by Herrmann, Hunter \& Elmegreen (2013).  But these features are rare in LSB
galaxies (which may explain the lack of double exponentials in our sample).

The most likely mechanism is the one proposed by Elmegreen \& Struck (2013), where
stellar scattering off of baryon clumps (stars or gas) lead to the formation of
exponential profiles.  In their simulations, a sufficiently strong irregular
morphology, which very much describes an LSB sample, drives a uniform and cold disk
into an exponential shape.  For objects in our sample, where most of the baryons are
located in distinct regions, the timescale for an exponential formation is less than
a Gyr.  Their simulations also indicate that asymmetric features tend to be
permanent, which explains why the internal colors in LSB galaxies are so uniform
between the lower and higher surface brightness regions, indicating the stellar
populations across an LSB disk have evolved in lockstep with the surface brightness
of the feature (Schombert, McGaugh \& Maciel 2013).

\subsection{Optical to Mid-IR Luminosities/Colors}

Also shown in Figure \ref{scalelengths} is a comparison of total magnitudes in $V$
(or $g$) and 3.6$\mu$m.  Again the correspondence is excellent (a mean $V-3.6$ color
of 2.3 is assumed for the unity line), and the lack of a broader scatter is a
statement concerning the low range in $V-3.6$ color for late-type galaxies. LSB
galaxies display better uniformity in color at 3.6$\mu$m than the optical bandpasses.
Converting 3.6$\mu$m to stellar mass (McGaugh \& Schombert 2013) gives a range of
10$^{7}$ to 10$^{10} M_{\sun}$ for LSB galaxies with a mean of $10^{9} M_{\sun}$
versus the HSB galaxies in the sample with a mean of $5 \times 10^{10} M_{\sun}$.
These value are consistent with optical derived masses.

\begin{figure}[!ht]
\centering
\includegraphics[scale=0.80,angle=0]{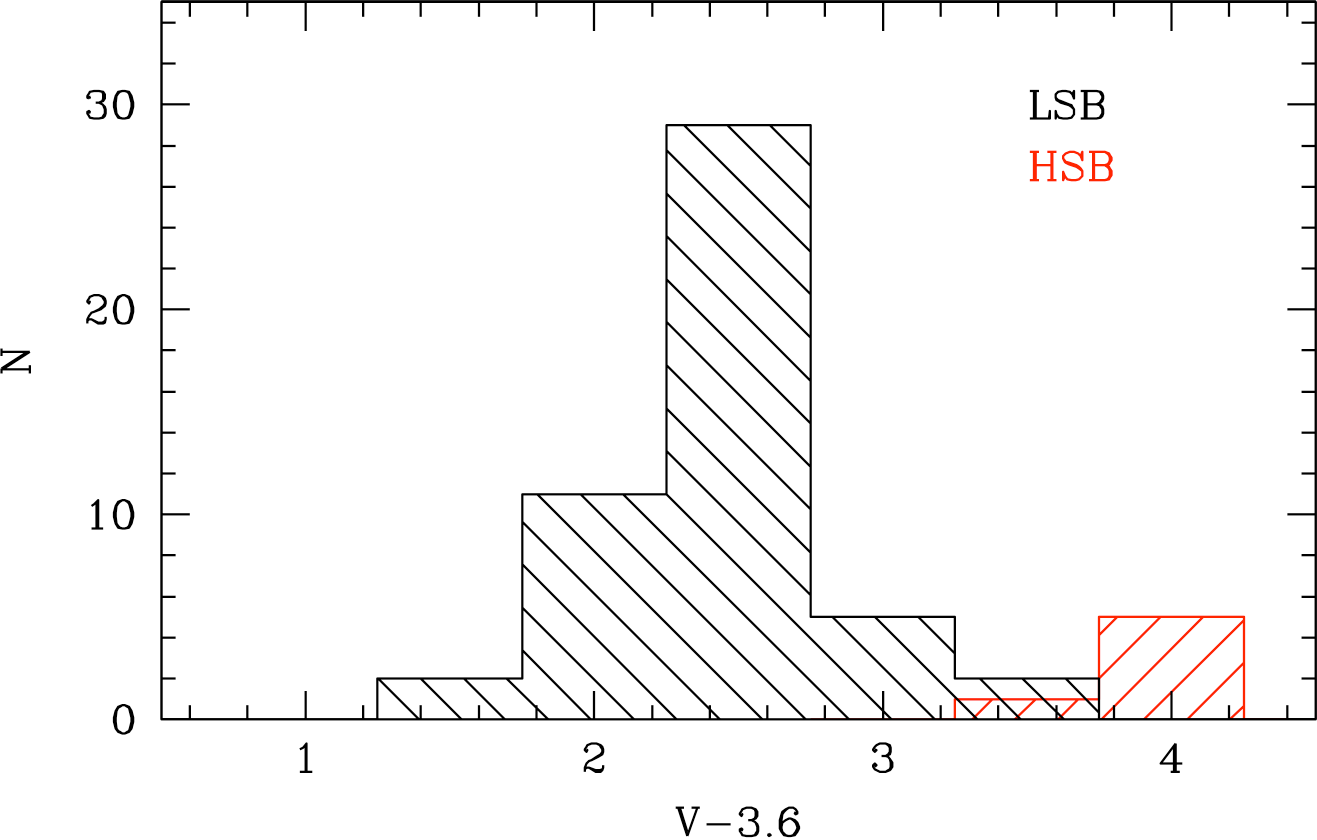}
\caption{\small Histogram of the total $V-3.6$ colors for the LSB and HSB galaxies in
our sample.  A mean color of 2.3 with a standard deviation of 0.5 is
measured for the sample.
}
\label{color_hist}
\end{figure}

Mean $V-3.6$ colors are shown in Figure \ref{color_hist} for both the LSB and HSB
galaxies in the sample with both optical and mid-IR imaging.  These are isophotal
colors, rather than aperture colors, meaning they are weighted by number of pixels
rather than luminosity of the pixels.  This produces a total color that underweights
bulge and core regions, and emphasizes the color of the LSB regions.  Despite the
different methods, the isophotal colors never differed from the total aperture colors
by more than 0.3 mags.

The mean $V-3.6$ color for the sample was 2.3, excluding the HSB galaxies from the
average.  As can be seen from Figure \ref{color_hist}, the HSB galaxies are 1.5 mags
redder than the LSB galaxies although we note that the LSB sample has a mean absolute
magnitude of $-$19 and the HSB galaxies have a mean of $-$24, meaning that some of
this color difference could be due to the mass-metallicity effect.  However, the SFR
is higher in HSB galaxies which should offset any metallicity effects.

The canonical test of the mass-metallicity effect is the color-magnitude relation
(CMR, Tully \etal 1982; Peletier \& de Grijs 1998).  Although the CMR is clearer for
early-type galaxies (whose lack of current star formation produces a relationship
dominated solely by metallicity effects), the CMR for spirals and irregulars is also a
useful stellar population diagnostic.  For example, regardless of the dominant
processes (star formation or chemical evolution), comparison of the CMR in spirals
and irregulars between clusters and the fields illuminates environmental processes
(Mobasher \etal 1986).

\begin{figure}[!ht]
\centering
\includegraphics[scale=0.80,angle=0]{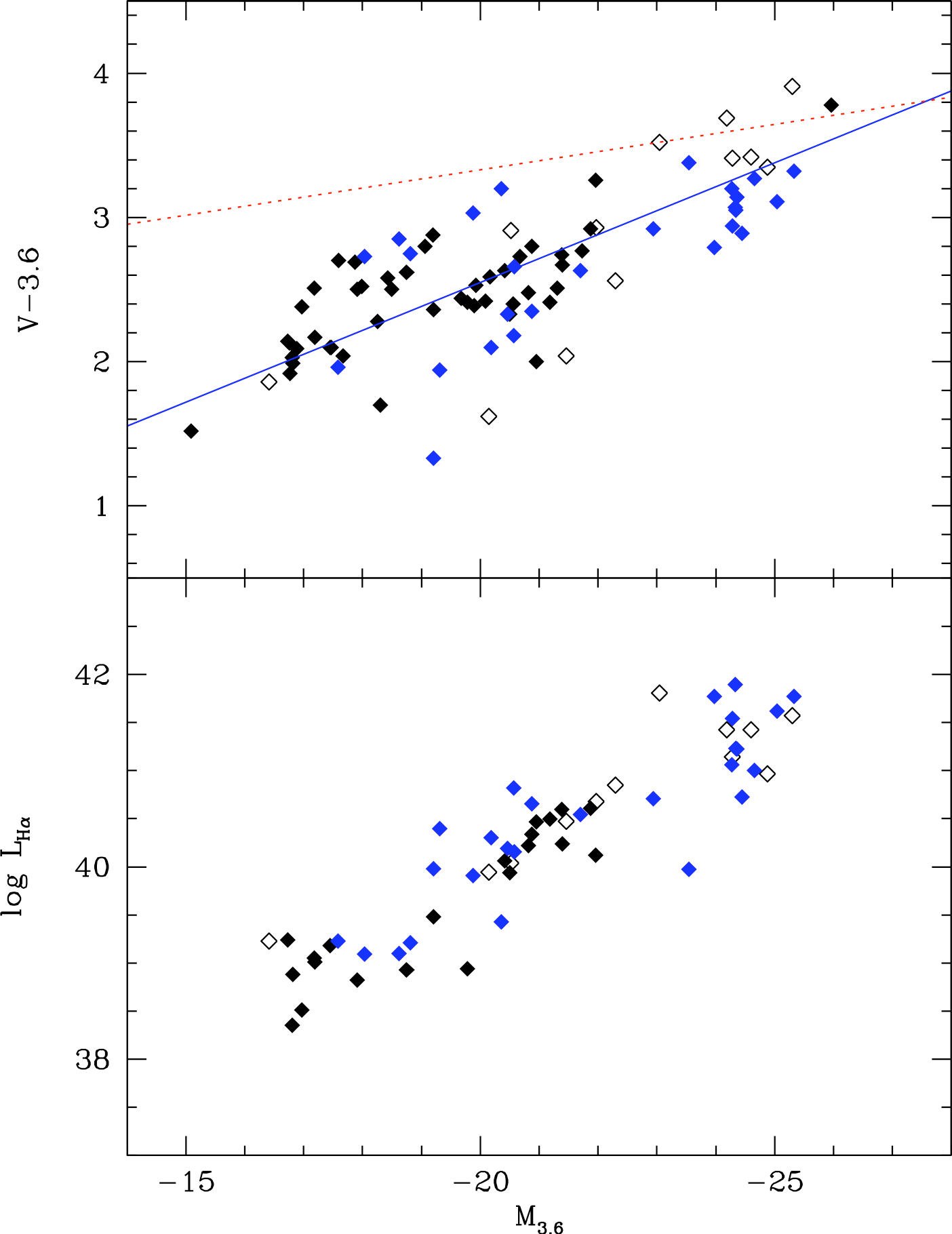}
\caption{\small The top panel displays the color-magnitude relation (CMR) for our
sample (black symbols, solid for LSB, open for HSB) and the Dale \etal sample of
early-type spirals (blue).  The relationship for ellipticals and S0's (Falcon-Barroso
\etal 2011) as the red line, the blue line is the fit to spiral galaxies from
Peletier \& de Grijs (1998) corrected to V-3.6 using a mean colors of $K-3.6=0.3$.
The steeper slope, compared to ellipticals, indicates that star formation/age is an
increasingly dominant component to the CMR over pure metallicity effects.  The bottom
panel displays the relationship between H$\alpha$ luminosity (i.e., current SFR) and
stellar mass (absolute 3.6$\mu$m luminosity).  
}
\label{cmr}
\end{figure}

The CMR for our sample is shown in the top panel of Figure \ref{cmr}.  The various
symbols represent our sample, the THINGS data (Leroy \etal 2008) and the sample from
Dale \etal (2005), all of which cover a range of luminosities and Hubble types.   The
red line is the CMR for ellipticals, which is pure mass-metallicity.  No corrections
are made for internal extinction.  While dust is minimal in LSB galaxies, so of the
scatter in the THINGS and Dale datasets are due to the lack of extinction
corrections.  Color gradients are also quite strong in spiral galaxies (see \S3.3), so
this effect will also contribute to the large scatter around a linear best fit.

Despite the large scatter, the correlation between color and luminosity is clear in
Figure \ref{cmr}.  Brighter galaxies tend to be earlier in Hubble type, thus, the
relationship converges onto the elliptical sequence as the bulge light dominates a
galaxy's color.  However, the slope is much steeper than the elliptical/S0 sequence,
therefore metallicity can not be the sole component to the late-type CMR (at least,
not global metallicity set by the onset of galactic winds).  Stellar mass (using
$M_{3.6}$ as a proxy) is also strongly correlated with H$\alpha$ luminosity (a proxy
for the SFR, see bottom panel of Figure \ref{cmr}).  However, the relationship
between SFR and color is inverse to the expected bluer colors with more star
formation.

The solution appears to be a combination of the explanations proposed by Tully, Mould
\& Aaronson (1982) and Peletier \& de Grijs (1998).  First, the spiral CMR is due, in
part, to an increase in the ratio of young to old stellar populations for later type
galaxies.  Evolution from a star-forming spiral to an S0 (by mass) must begin with an
abrupt cessation of star formation to jump from the spiral CMR to the elliptical/S0
CMR (crossing the `green valley', Strateva \etal 2001).  The analysis by Peletier \&
de Grijs shows that decreasing mean age is insufficient to explain the spiral CMR
slope, and an additional metallicity component is required.  This agrees well with
the studies of [OIII] lines in dwarf galaxies (Zahid \etal 2012) and large color
surveys with SDSS (Tojeiro \etal 2013) where galaxies of the same morphology have
their colors correlated with SFR, not mass.

\begin{figure}[!ht]
\centering
\includegraphics[scale=0.80,angle=0]{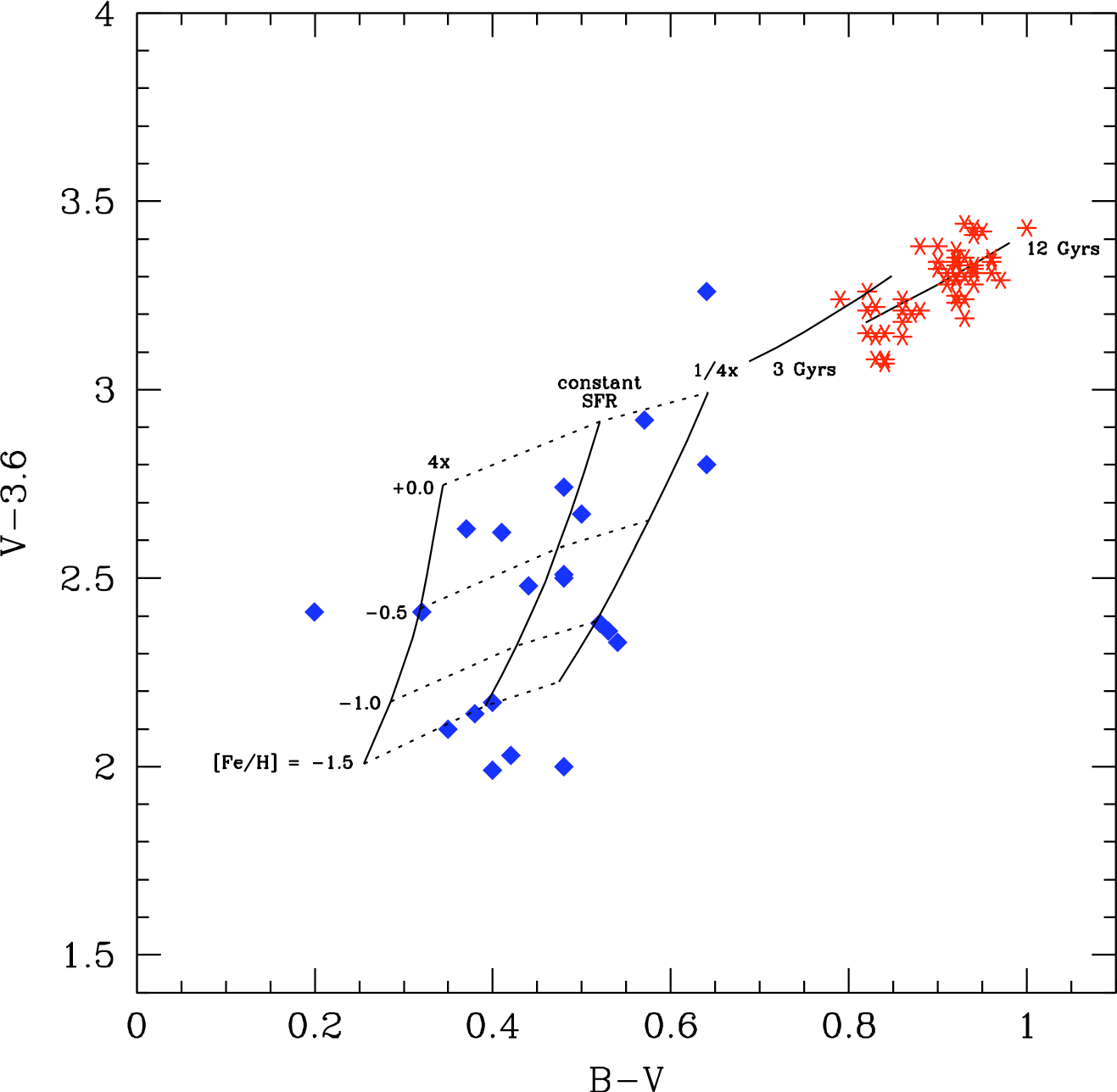}
\caption{\small The two color diagram, $B-V$ versus $V-3.6$, for the LSB galaxies in
our sample with both optical and mid-IR photometry.  The red symbols are early-type
galaxies from Falcon-Barroso \etal (2011), blue symbols are LSB galaxies.  The two
solid lines are 12 and 3 Gyrs multi-metallicity burst models from Schombert \& Rakos
(2009) based on Bruzual \& Charlot (2003) isochrones.  The grid represents models of
constant star formation over 12 Gyrs with varying terminal [Fe/H] values.  Bluer and
redder models are constructed by assuming a four-fold increase/decrease in star
formation over the last 0.5 Gyrs.
}
\label{pop}
\end{figure}

Further illumination to the underlying stellar population in LSB galaxies is given by
the two color $B-V$ versus $V-3.6$ diagram, shown in Figure \ref{pop}.  The red
symbols are early-type galaxies from Falcon-Barroso \etal (2011), the blue symbols
are the LSB galaxies from our sample.  Also shown are single burst 12 and 3 Gyrs
multi-metallicity models (Rakos \& Schombert 2009) where the red side of the models
represents [Fe/H]=+0.3 and the blue side represents [Fe/H]=$-$0.5.  A majority of
ellipticals are well described by a single age of 12 Gyrs with the variation in color
due to the mass-metallicity effect.  S0 and Sa type galaxies have slightly younger
mean ages, but none are bluer than the 3 Gyrs model.

Star-forming galaxies (late-type) display much bluer optical and mid-IR colors, and a
much larger range in $V-3.6$ color.  The range of color are incompatible with any
single age or frosting model (where an old population contains a small fraction of
young stars, Schombert \& McGaugh 2013).  We note that the current star formation
rate of LSB galaxies, divided by their stellar mass, is near a Hubble time
(Schombert, McGaugh \& Maciel 2013).  This implies that the mean star formation rate
in LSB galaxies has been nearly constant for their lifetimes.  While this rate may
vary wildly at any particular epoch, a first order stellar population model is one
which assumes a constant star formation rate then sums the luminosities over all the
generations to form a final color.  It is this type of model that is shown in Figure
\ref{pop}.

The constant star formation model shown in Figure \ref{pop} is based on modified
Bruzual \& Charlot SSP's (single stellar populations, single age and metallicity
isochrones).  The BC03 models were successfully modified to reproduce the CMR for
ellipticals using a multi-metallicity technique (Odell, Schombert \& Rakos 2002).
Multi-metallicity populations are generated by selecting BC03 models of a single age
and summing various metallicities using the shape of the metallicity distribution in
our Galaxy, and allowing the peak to vary to represent a changing mean [Fe/H].  To
model a constant star formation population, the total population is divided into a
number of bins from 12 Gyrs to the present.  Each age is assigned a mean [Fe/H]
starting at [Fe/H]=$-1.7$ at 12 Gyrs and advanced to a final value that varied from
$-$1.5 to +0.3 following a chemical evolution prescription (Prantzos 2009) and
following the CMR with redshift (Zahid \etal 2013).  Each bin is summed to and
weighted by the mean luminosity compared to the 12 Gyr bin (i.e., older, more
metal-poor stars tend to be brighter than their younger, metal-rich counterparts).

In addition, as our data covers the mid-IR portion, we extended the BC03 models to
cover TP-AGB evolutionary tracks (Marigo \etal 2007).  This correction begins for
populations older than 40 Myrs with a $\Delta(V-3.6)$=0.5, dropping to 0.2 by 500 Myrs
and less than 0.05 for greater than 2 Gyrs.  While this correction is applied to each
age bin, the total effect is $\Delta(V-3.6)$=0.25 with respect to models without the
AGB correction.  This is a significant difference in Figure \ref{pop} resulting in
primarily overestimating an LSB galaxy's metallicity.  We have ignored extinction
effects due to the low dust content in LSB galaxies (these models are described in
greater detail in Schombert \& McGaugh 2013).

The mid line in Figure \ref{pop} presents the final model with a run of mean galaxy
metallicity from $<Fe/H>=-1.5$ to solar.  Due to the stochastic nature of star
formation, we also considered two other models where the star formation was suppressed
by a factor of four for the last 0.5 Gyrs and where the star formation was enhanced by a
factor of four for the last 0.5 Gyrs (4x and 1/4x).  The resulting model tracks are shown
in Figure \ref{pop} where changing the recent star formation has the largest effect
in $B-V$ color as expected.

A majority of the LSB colors fall within the range of models described above.  The
mean color for the sample falls exactly on the midline model, although the color errors
prevent any exact mapping of star formation history to a particular galaxy.  We can
rule out a majority of the stars in LSB galaxies being formed in the last few Gyrs,
or their $V-3.6$ colors would be much redder to due a strong TP-AGB component.  The
spread in colors are well matched by the models with enhanced/suppressed SFR by
factors of four, which is consistent with the spread in H$\alpha$ emission as a
function of stellar mass for LSB galaxies (Schombert, McGaugh \& Maciel 2013).

\subsection{Color Gradients}

\begin{figure}[!ht]
\centering
\includegraphics[scale=0.80,angle=0]{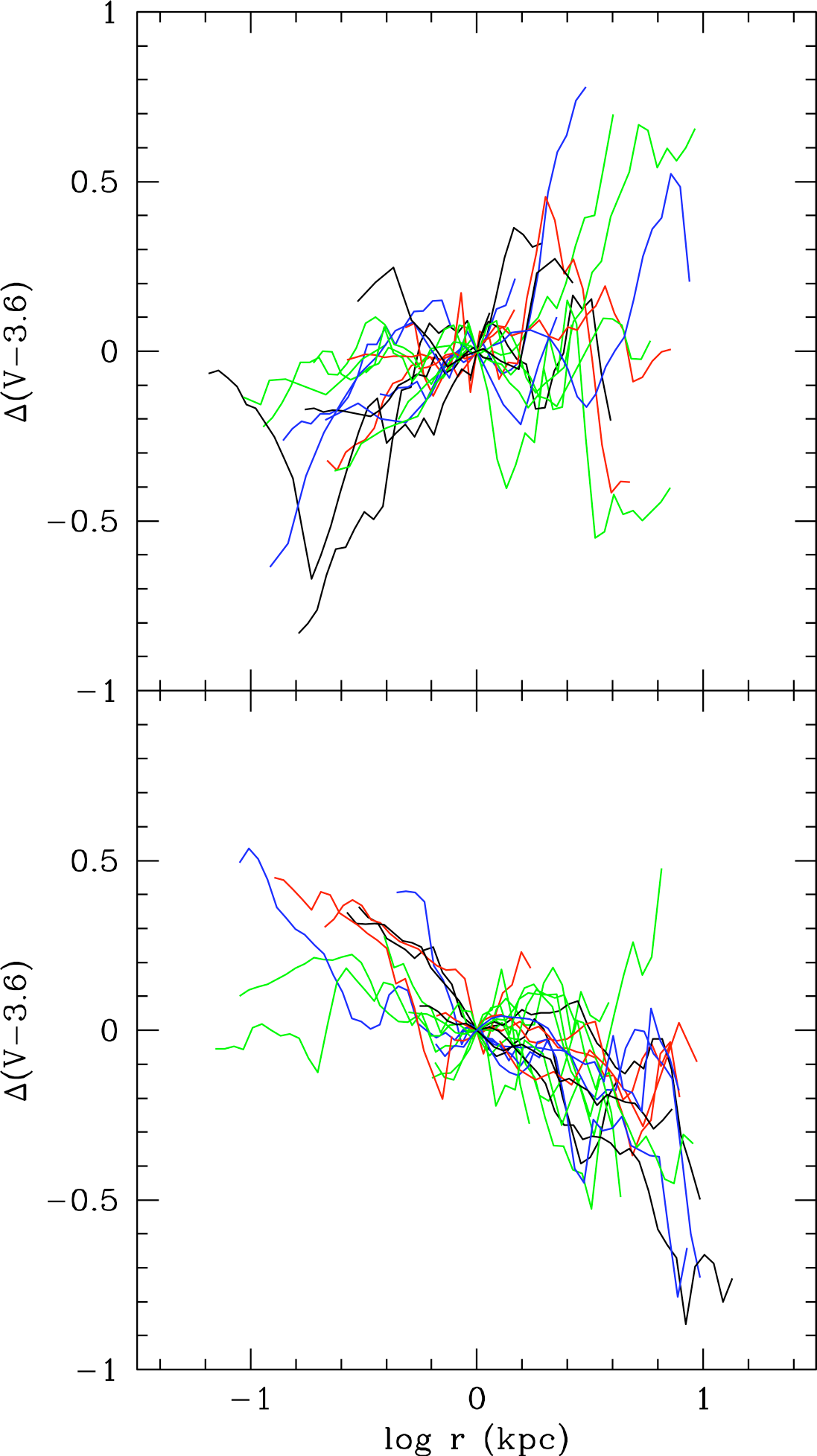}
\caption{\small The $V-3.6$ color gradients for 45 LSB galaxies.  The gradients are
normalized to $r=1$ kpc.  Flat or rising gradients are shown in the top panel,
downward gradients are displayed in the bottom panel.  The sample is evenly divided
into flat/upward versus downward gradients.  LSB galaxies with late-type morphology or
low baryon mass tend to have flat or upward gradients.
}
\label{gradients}
\end{figure}

Color gradients between $V$ (or $g$) and 3.6$\mu$m are found at the bottom of each
surface brightness plot.  These are calculated by interpolating the 3.6$\mu$m surface
brightness value at each $V$ surface brightness point.  The resulting color gradients
are shown in Figure \ref{gradients} where the color profiles have be normalized to $r
= 1$ kpc and divided into two groups, 1) those with flat or slightly rising profiles
and 2) those with decreasing colors with radius.  The sample divides evenly into
those galaxies with flat versus decreasing color gradients.  The
mean down gradient is $\Delta(V-3.6)$/log $R = -$0.4, the mean up gradient is 
$\Delta(V-3.6)$/log $R = +$0.3.

Galaxies with decreasing gradients are predominately disk-like in morphology.  Those
galaxies with flat gradients are irregular in morphology or dwarf-like.  Galaxies
with decreasing gradients are brighter, on average, than flat or increasing gradient
galaxies, but both types are found at all masses.  There is a sharp transition with
respect to baryon mass.  A majority of the galaxies with baryon masses less than
$10^9 M_{\sun}$ have flat or upward color gradients, those galaxies with masses
greater than $10^9 M_{\sun}$ have downward gradients.  This is similar to the result
found by Tortora \etal (2010) for 50,000 SDSS galaxies.  

While color gradients in early-type galaxies are primary due to metallicity effects,
disk and late-type galaxies have significant star formation with could produce a
color gradient based on mean local stellar age.  Since the CMR for our our sample is
significantly steeper than the slope of the CMR for ellipticals, star formation must
play a strong role in the observed color gradients herein.  In addition, those
galaxies with downward gradients have, on average, 100 times more H$\alpha$ emission
than flat gradient galaxies in our sample.  Flat gradients for low mass LSB galaxies
probably reflects their already low mean metallicities, due to a suppressed past SFR,
such that there is very little dynamic range in [Fe/H] within the entire galaxy.

All the galaxies with up or flat gradients have no gradients in $B-V$.  And a
majority of the galaxies with downward gradients also have no or very modest $B-V$
gradients.  The lack of optical gradients is a known property of LSB galaxies
(Schombert, McGaugh \& Maciel 2013), mostly due to the fact that star formation is
more dispersed in LSB galaxies compared to their HSB cousins.  This implies that the
LSB gradients are dominated by metallicity effects, rather than age (recent SF, see
Figure \ref{pop}) and the low SFR in LSB galaxies may allow a sharper analysis of
metallicity gradients (i.e., spatial chemical evolution) due to a decoupling of SF
effects.  

\section{Summary}

Deep {\it Spitzer} surface photometry is presented for 61 LSB galaxies (45
with matching optical imaging) and an additional 14 HSB galaxies.  {\it
Spitzer} imaging has several unique data reduction challenges, such as a
sharp increase in the number of background point sources, all of which
can be resolved with well designed software routines.  The data presented
herein have comparable depth to sky-limited optical imaging, but at
wavelengths that are impossible to achieve from the ground and measure the
portion of a galaxy's SED critical to estimates of galaxy stellar mass and
star formation history.  Our results are summarized as the following:

\begin{description}

\item{(1)} Our sample, selected for low central surface brightness ($\mu_o
< 19$ mag arcsecs$^{-2}$ at 3.6$\mu$m), covers a luminosity range of $-$11
to $-$22 in $M_{3.6}$, a size range of 1 to 25 kpc and gas mass fraction of 30
to 95\%.  While a majority of the sample is gas-rich ($f_g > 0.5$), the
only defining characteristic to the sample is central surface brightness,
not mass or size.

\item{(2)} Surface brightness profiles at 3.6$\mu$m have identical features
to profiles at $V$, regardless of galaxy morphology.  Contrary to studies
of HSB dwarfs (Herrmann, Hunter \& Elmegreen 2013), we find a majority of
LSB galaxies have single exponential profiles with only 15\% displaying
upward or downward breaks from a single exponential.  Part of this
difference may be due to the lower S/N in LSB envelopes which disguises
outer changes in the profile shape.

\item{(3)} $V-3.6$ colors of LSB galaxies are bluer than their HSB
counterparts with a mean color of 2.3 and a standard deviation of 0.5.  The
color-magnitude relation (CMR) is evident for LSB galaxies, with a similar
slope to that found for normal spirals, indicating that star formation
(age) dominates over metallicity effects.

\item{(4)} The timescale for total star formation in LSB galaxies
($M_*$/SFR, the ratio of the total stellar mass to the current SFR) are
near a Hubble time implying a history of constant, albeit extremely low,
star formation. This is consistent with the fact that single age stellar
population models (where all the stars are of a single age, but varying
metallicities) fail to reproduce the blue optical colors compared with red
near-IR colors (a signature of a mixed age stellar population).  However,
the mean LSB optical to mid-IR colors do compare well with models of
constant star formation, as can be seen in Figure \ref{pop}.

\item{(5)} The spread in $B-V$ color for LSB galaxies can be modeled by
assuming a fourfold decrease/increase in star formation over the last 0.5
Gyrs.  These quasi-stochastic bursts of star formation would explain the
weak correlation between SFR and surface brightness (Paper I, where low SFR
is associated with low stellar density galaxies) such that the wide
variation in SFR versus surface brightness is due to the present SFR being
a snapshot of the current epoch and the mean surface brightness represents
the integrated effect of star formation over the age of the galaxy.

\item{(6)} Downward color gradients exist for 1/2 the sample, but low mass
and late-type morphology LSB galaxies display no gradients.  Gradients
appear to be driven more by metallicity effects as irregular LSB galaxies
have very low SFR's and are relatively uniform in disk color.

\end{description}

Our sample emphasizes the promise that mid-IR imaging has for studies
attempting to measure the total luminosity and size of a galaxy population,
particularly with respect to formation correlations such as the
Tully-Fisher relation and the Fundamental Plane.  More surprising is the
information provided by optical to mid-IR colors with respect to the
underlying stellar population.  The combination of near-IR colors with
current SFRs from H$\alpha$ imaging yields a clearer picture of star formation in
LSB galaxies than afforded by optical colors alone.  The model of roughly
constant mean star formation rate punctuated by stochastic variations in
current SFR agrees well with constraints from kinematic studies, providing
a natural explanation for the observed range of stellar mass-to-light
ratios.  We will use these as inputs towards a coherent history of the
evolution of LSB galaxies in the next paper of our series.

\acknowledgements This work is based on observations made with the {\it Spitzer}
Space Telescope, which is operated by the Jet Propulsion Laboratory, California
Institute of Technology under a contract with NASA. Support for this work was
provided by NASA through an award issued by JPL/Caltech.  The software for this
project was supported by NASA's AISR and ADP programs.

\pagebreak

\clearpage

\begin{deluxetable}{lccccccccccccc}
\tablecolumns{14}
\small
\tablewidth{0pt}
\tablecaption{LSB Sample Morphology}
\tablehead{
\colhead{Object} & \colhead{profile} & \colhead{Hubble} & \colhead{disk} & 
\colhead{optical} & \colhead{H$\alpha$} & \colhead{D} \\
\colhead{} & \colhead{type} & \colhead{type} & \colhead{type} & \colhead{source} & \colhead{} & \colhead{(Mpc)} \\
\colhead{(1)} & \colhead{(2)} & \colhead{(3)} & \colhead{(4)} & \colhead{(5)} & \colhead{(6)} & \colhead{(7)} \\
}
\startdata

CamB & I & Irr & X & $--$ & $--$ & 0.3 \\
D500-2 & I & Sm & D & MDM & $--$ & 21.9 \\
D500-3 & I & dI & X & KPNO & \checkmark & 22.7 \\
D500-4 & I & dI & X & SDSS & $--$ & 26.0 \\
D512-2 & I & dI & X & SDSS & $--$ & 14.1 \\
D564-8 & I & dI & D & MDM & $--$ & 10.4 \\
D568-2 & II & Im & D & KPNO & \checkmark & 21.2 \\
D570-4 & III: & Im & D & MDM & $--$ & 19.1 \\
D572-5 & III & Irr & X & KPNO & \checkmark & 18.0 \\
D575-1 & III & Im & D & SDSS & $--$ & 12.0 \\
D575-7 & II & Im & X & KPNO & \checkmark & 18.0 \\
D584-2 & I & Im & B+D & SDSS & $--$ & 37.5 \\
D631-7 & II & dI & X & KPNO & \checkmark & 7.8 \\
D640-13 & I & dI: & D & SDSS & $--$ & 19.0 \\
D723-5 & I & Sd & D & KPNO & \checkmark & 27.7 \\
DDO064 & I & Im & X & SDSS & $--$ & 10.8 \\
DDO168 & I & Im & D & KPNO & \checkmark & 5.2 \\
DDO170 & I & Im & X & SDSS & $--$ & 16.5 \\
ESO215-G009 & II & SAB(s)m: & X & $--$ & $--$ & 12.0 \\
F415-3 & I & Sm & D & KPNO & \checkmark & 10.4 \\
F530-3 & I & Sc & B+D & $--$ & $--$ & 61.7 \\
F561-1 & I: & Sm & D & KPNO & \checkmark & 69.8 \\
F563-1 & I & Sm/Im & B+D & KPNO & \checkmark & 52.2 \\
F563-V1 & I & dI & D & KPNO & \checkmark & 57.6 \\
F563-V2 & I & Irr & D & SDSS & $--$ & 63.5 \\
F565-V2 & I & Im & D & KPNO & \checkmark & 55.1 \\
F567-2 & I & Sm & D & SDSS & $--$ & 83.4 \\
F568-1 & I & Sc & B+D & KPNO & \checkmark & 95.4 \\
F568-3 & I & Sd & D & KPNO & \checkmark & 86.7 \\
F568-6 & I & S/Malin-like & B+D & SDSS & $--$ & 201.0 \\
F568-V1 & I & S/Malin-like & B+D & KPNO & \checkmark & 84.8 \\
F571-5 & I & Sm & D & SDSS & $--$ & 63.5 \\
F571-8 & I & Sb & B+D & SDSS & $--$ & 56.2 \\
F571-V1 & I & Sd/Sm & D & SDSS & $--$ & 84.3 \\
F574-1 & I & Sd & B+D & KPNO & \checkmark & 100.0 \\
F574-2 & I & Sm: & D & KPNO & \checkmark & 92.2 \\
F577-V1 & I & Sd: & D & KPNO & \checkmark & 113.0 \\
F579-1 & I & Sb & B+D & SDSS & $--$ & 90.5 \\
F583-1 & I & Sm/Irr & D & SDSS & $--$ & 32.5 \\
F583-5 & I & Sb & B+D & SDSS & $--$ & 46.7 \\
F585-3 & I & Sm & B+D: & SDSS & $--$ & 43.7 \\
F585-V1 & I & dI & D & SDSS & $--$ & 28.1 \\
F611-1 & I & dI/Im & D & KPNO & \checkmark & 25.5 \\
F746-1 & I & Irr & D & KPNO & \checkmark & 105.0 \\
GR8 & II & ImV & X & $--$ & $--$ & 7.2 \\
KK98-251 & I & Irr? & D & $--$ & $--$ & 5.5 \\
N3741 & I & ImIII/BCD & B+D & SDSS & $--$ & 6.2 \\
U128 & I & Sdm & B+D & KPNO & \checkmark & 58.4 \\
U628 & I & Sm: & D & $--$ & $--$ & 71.1 \\
U1195 & I & Sc(f) & B+D & SDSS & $--$ & 6.6 \\
U1230 & I & Sm: & B+D & $--$ & $--$ & 49.2 \\
U2014 & II & Im: & X & $--$ & $--$ & 4.8 \\
U5005 & I & Im & D & KPNO & \checkmark & 57.0 \\
U5209 & I & Im & D & SDSS & $--$ & 11.0 \\
U5750 & I: & SBdm: & D & SDSS & $--$ & 62.2 \\
U5999 & I & Im & D & $--$ & $--$ & 51.9 \\
U11557 & I & SAB(s)dm & B+D & SDSS & $--$ & 16.7 \\
U12082 & I & Sm & B+D & $--$ & $--$ & 6.5 \\
U12212 & I & Sm: & D & $--$ & $--$ & 7.5 \\
U12632 & I & Sm: & B+D & $--$ & $--$ & 1.6 \\
U12695 & I & Sm: & D & $--$ & $--$ & 80.9 \\

\enddata
\tablecomments{(1) Galaxy name. (2) surface brightness profile type,
Erwin \etal (2008). (3) Hubble type from Schombert \etal
(1992,1997). (4) surface brightness profile structure, D = disk, B+D = bulge+disk, X
= unknown. (5) source of optical data. (6) H$\alpha$ images available. (7) distance
in Mpc from Tully \etal (2009).
}
\end{deluxetable}

\begin{deluxetable}{lccccccccccccc}
\tablecolumns{14}
\small
\tablewidth{0pt}
\tablecaption{HSB Sample Morphology}
\tablehead{
\colhead{Object} & \colhead{profile} & \colhead{Hubble} & \colhead{disk} & 
\colhead{optical} & \colhead{H$\alpha$} & \colhead{D} \\
\colhead{} & \colhead{type} & \colhead{type} & \colhead{type} & \colhead{source} & \colhead{} & \colhead{(Mpc)} \\
\colhead{(1)} & \colhead{(2)} & \colhead{(3)} & \colhead{(4)} & \colhead{(5)} & \colhead{(6)} & \colhead{(7)} \\
}
\startdata

ESO563-G021 & II & SAbc: & B+D & $--$ & $--$ & 67.6 \\
N801 & I & Sc & B+D & $--$ & $--$ & 75.3 \\
N1003 & I & SA(s)cd & B+D & $--$ & $--$ & 10.2 \\
N1167 & I & SA0-;LINER & B+D & $--$ & $--$ & 66.0 \\
N2998 & I & SAB(rs)c & B+D & SDSS & $--$ & 68.3 \\
N5533 & I & SA(rs)ab & B+D & SDSS & $--$ & 59.4 \\
N6195 & I & Sb & B+D & SDSS & $--$ & 127.0 \\
N6674 & I & SB(r)b & B+D & $--$ & $--$ & 51.9 \\
U1551 & I & Sdm & B+D & $--$ & $--$ & 33.2 \\
U2259 & II & SB(s)dm & B+D & $--$ & $--$ & 5.3 \\
U2885 & I & SA(rs)c & B+D & $--$ & $--$ & 78.9 \\
U2953 & I & SA(s)ab & B+D & $--$ & $--$ & 11.2 \\
U5709 & I & Sd: & B+D & SDSS & $--$ & 91.0 \\
U11455 & I & Sc & B+D & $--$ & $--$ & 73.5 \\

\enddata
\end{deluxetable}

\begin{deluxetable}{lcccccccc}
\tablecolumns{14}
\small
\tablewidth{0pt}
\tablecaption{LSB Photometric Properties}
\tablehead{
\colhead{Object} & \colhead{$m_{3.6}$} & \colhead{$V-3.6$} & \colhead{$R_{Holmberg}$} & \colhead{$\mu_o$} & 
\colhead{log $M_*$} & \colhead{log $M_{gas}$} & \colhead{log $M_{baryons}$} & \colhead{$f_g$} \\
\colhead{} & \colhead{} & \colhead{(kpc)} & \colhead{(kpc)} & \colhead{($M_{\sun}$)} & \colhead{($M_{\sun}$)} & \colhead{($M_{\sun}$)}
& \colhead{} \\
\colhead{(1)} & \colhead{(2)} & \colhead{(3)} & \colhead{(4)} & \colhead{(5)} & \colhead{(6)} & \colhead{(7)} & \colhead{(8)} \\
}
\startdata

       CamB & 11.13 $\pm$ 0.03 &  $--$  &  0.2 & 20.26 &  5.50 &  5.19 &  5.67 & 0.33 \\
     D500-2 & 12.96 $\pm$ 0.07 &  1.70$\pm$0.19  &  3.8 & 19.92 &  8.49 &  9.20 &  9.28 & 0.84 \\
     D500-3 & 14.16 $\pm$ 0.12 &  2.51$\pm$0.01  &  3.9 & 19.54 &  8.04 &  8.25 &  8.46 & 0.62 \\
     D500-4 & 14.08 $\pm$ 0.12 &  2.52$\pm$0.23  &  2.7 & 19.58 &  8.19 &  7.94 &  8.38 & 0.36 \\
     D512-2 & 12.88 $\pm$ 0.07 &  2.69$\pm$0.28  &  3.8 & 19.88 &  8.14 &  8.20 &  8.47 & 0.53 \\
     D564-8 & 13.98 $\pm$ 0.11 &  1.52$\pm$0.11  &  1.3 & 21.22 &  7.44 &  7.81 &  7.97 & 0.70 \\
     D568-2 & 14.82 $\pm$ 0.16 &  2.03$\pm$0.18  &  1.7 & 19.78 &  7.72 &  7.94 &  8.15 & 0.63 \\
     D570-4 & 13.74 $\pm$ 0.10 &  2.04$\pm$0.11  &  2.7 & 20.47 &  8.06 &  8.35 &  8.53 & 0.66 \\
     D572-5 & 14.01 $\pm$ 0.11 &  1.99$\pm$0.07  &  2.2 & 19.75 &  7.90 &  8.37 &  8.50 & 0.75 \\
     D575-1 & 12.48 $\pm$ 0.06 &  2.10$\pm$0.16  &  4.3 & 20.95 &  8.16 &  8.06 &  8.41 & 0.44 \\
     D575-7 & 14.09 $\pm$ 0.12 &  2.17$\pm$0.11  &  2.1 & 19.72 &  7.87 &  8.42 &  8.53 & 0.78 \\
     D584-2 & 13.67 $\pm$ 0.10 &  2.88$\pm$0.10  &  6.5 & 20.17 &  8.67 &  8.98 &  9.15 & 0.67 \\
     D631-7 & 11.98 $\pm$ 0.05 &  2.14$\pm$0.09  &  2.4 & 19.66 &  7.99 &  8.48 &  8.61 & 0.76 \\
    D640-13 & 13.87 $\pm$ 0.11 &  1.92$\pm$0.15  &  2.4 & 20.60 &  8.00 &  8.16 &  8.39 & 0.59 \\
     D723-5 & 13.00 $\pm$ 0.07 &  2.36$\pm$0.20  &  4.6 & 19.66 &  8.68 &  8.43 &  8.87 & 0.36 \\
     DDO064 & 11.91 $\pm$ 0.04 &  2.28$\pm$0.09  &  4.0 & 19.36 &  8.30 &  8.85 &  8.96 & 0.78 \\
     DDO168 & 10.69 $\pm$ 0.03 &  2.10$\pm$0.06  &  3.3 & 19.62 &  8.15 &  8.78 &  8.87 & 0.81 \\
     DDO170 & 12.56 $\pm$ 0.06 &  2.50$\pm$0.05  &  7.2 & 19.54 &  8.41 &  9.03 &  9.12 & 0.81 \\
ESO215-G009 & 11.87 $\pm$ 0.04 &  $--$  &  3.4 & 18.95 &  8.41 &  9.76 &  9.78 & 0.96 \\
     F415-3 & 13.12 $\pm$ 0.08 &  2.38$\pm$0.22  &  2.8 & 20.68 &  7.78 &  8.65 &  8.71 & 0.88 \\
     F530-3 & 13.06 $\pm$ 0.07 &  $--$  &  9.6 & 19.71 &  9.35 &  9.55 &  9.76 & 0.61 \\
     F561-1 & 13.34 $\pm$ 0.08 &  2.80$\pm$0.04  & 10.1 & 19.89 &  9.35 &  9.26 &  9.61 & 0.45 \\
     F563-1 & 13.18 $\pm$ 0.08 &  2.63$\pm$0.11  & 10.5 & 20.78 &  9.16 &  9.66 &  9.78 & 0.76 \\
    F563-V1 & 14.02 $\pm$ 0.11 &  2.41$\pm$0.10  & 10.6 & 20.79 &  8.91 &  8.94 &  9.23 & 0.52 \\
    F563-V2 & 13.45 $\pm$ 0.09 &  2.40$\pm$0.15  &  9.5 & 19.40 &  9.22 &  9.54 &  9.71 & 0.68 \\
    F565-V2 & 14.96 $\pm$ 0.17 &  2.62$\pm$0.18  &  6.3 & 20.80 &  8.50 &  8.99 &  9.11 & 0.76 \\
     F567-2 & 14.44 $\pm$ 0.14 &  2.59$\pm$0.15  &  8.4 & 20.64 &  9.06 &  9.39 &  9.56 & 0.68 \\
     F568-1 & 13.51 $\pm$ 0.09 &  2.67$\pm$0.06  & 14.7 & 20.41 &  9.55 &  9.72 &  9.94 & 0.60 \\
     F568-3 & 12.96 $\pm$ 0.07 &  2.77$\pm$0.06  & 13.5 & 18.89 &  9.69 &  9.62 &  9.96 & 0.46 \\
     F568-6 & 10.56 $\pm$ 0.02 &  3.78$\pm$0.19  & 77.5 & 20.22 & 11.38 & 10.67 & 11.46 & 0.16 \\
    F568-V1 & 13.82 $\pm$ 0.10 &  2.48$\pm$0.03  &  9.6 & 19.92 &  9.32 &  9.70 &  9.86 & 0.71 \\
     F571-5 & 14.33 $\pm$ 0.13 &  2.44$\pm$0.11  &  8.1 & 20.52 &  8.87 &  9.49 &  9.58 & 0.81 \\
     F571-8 & 11.86 $\pm$ 0.04 &  4.09$\pm$0.30  & 12.2 & 19.96 &  9.75 &  9.33 &  9.89 & 0.27 \\
    F571-V1 & 14.54 $\pm$ 0.14 &  2.42$\pm$0.12  &  7.3 & 20.29 &  9.03 &  9.30 &  9.49 & 0.65 \\
     F574-1 & 13.62 $\pm$ 0.10 &  2.74$\pm$0.20  & 15.2 & 19.55 &  9.55 &  9.73 &  9.95 & 0.60 \\
     F574-2 & 14.32 $\pm$ 0.13 &  2.33$\pm$0.13  & 10.8 & 20.77 &  9.19 &  9.29 &  9.54 & 0.55 \\
    F577-V1 & 14.32 $\pm$ 0.13 &  2.00$\pm$0.09  & 12.0 & 20.26 &  9.37 &  9.78 &  9.92 & 0.72 \\
     F579-1 & 12.82 $\pm$ 0.07 &  $--$  & 13.1 & 19.05 &  9.78 &  9.49 &  9.96 & 0.34 \\
     F583-1 & 13.50 $\pm$ 0.09 &  2.80$\pm$0.27  &  6.0 & 20.35 &  8.62 &  9.39 &  9.46 & 0.86 \\
     F583-5 & 13.42 $\pm$ 0.09 &  2.53$\pm$0.27  &  9.5 & 21.91 &  8.97 &  9.62 &  9.71 & 0.82 \\
     F585-3 & 13.30 $\pm$ 0.08 &  2.39$\pm$0.22  &  9.0 & 20.48 &  8.95 &  9.70 &  9.77 & 0.85 \\
    F585-V1 & 14.65 $\pm$ 0.15 &  2.70$\pm$0.06  &  1.6 & 20.61 &  8.03 &  8.83 &  8.90 & 0.86 \\
     F611-1 & 14.12 $\pm$ 0.12 &  2.50$\pm$0.12  &  3.0 & 20.92 &  8.16 &  8.51 &  8.67 & 0.69 \\
     F746-1 & 13.80 $\pm$ 0.10 &  2.51$\pm$0.19  & 18.2 & 19.58 &  9.52 &  9.97 & 10.11 & 0.74 \\
        GR8 & 12.70 $\pm$ 0.06 &  $--$  &  1.5 & 20.90 &  7.63 &  8.17 &  8.28 & 0.78 \\
   KK98-251 & 12.53 $\pm$ 0.06 &  $--$  &  2.9 & 20.52 &  7.46 &  8.03 &  8.13 & 0.78 \\
      N3741 & 12.08 $\pm$ 0.01 &  2.09$\pm$0.07  &  1.6 & 19.47 &  7.75 &  8.78 &  8.82 & 0.92 \\
       U128 & 11.96 $\pm$ 0.05 &  2.92$\pm$0.10  & 18.1 & 20.26 &  9.51 &  9.98 & 10.10 & 0.75 \\
       U628 & 12.32 $\pm$ 0.01 &  $--$  & 15.0 & 19.08 &  9.77 &  9.85 & 10.11 & 0.54 \\
      U1195 & 10.66 $\pm$ 0.01 &  2.58$\pm$0.11  &  4.2 & 19.09 &  8.37 &  8.63 &  8.82 & 0.65 \\
      U1230 & 12.18 $\pm$ 0.05 &  $--$  & 12.0 & 20.21 &  7.70 &  7.81 &  8.06 & 0.56 \\
      U2014 & 12.53 $\pm$ 0.06 &  $--$  &  1.6 & 19.68 &  9.26 &  9.58 &  9.75 & 0.68 \\
      U5005 & 12.60 $\pm$ 0.06 &  2.41$\pm$0.08  & 11.9 & 20.00 &  9.33 &  9.03 &  9.51 & 0.33 \\
      U5209 & 13.45 $\pm$ 0.09 &  2.13$\pm$0.21  &  1.8 & 20.69 &  9.46 &  9.26 &  9.67 & 0.39 \\
      U5750 & 13.30 $\pm$ 0.08 &  2.73$\pm$0.27  & 12.9 & 19.50 &  8.26 &  8.58 &  8.75 & 0.67 \\
      U5999 & 12.74 $\pm$ 0.06 &  $--$  & 11.3 & 20.31 &  8.01 &  8.49 &  8.61 & 0.75 \\
     U11557 &  9.96 $\pm$ 0.02 &  3.60$\pm$0.16  &  7.9 & 18.13 &  7.24 &  7.76 &  7.88 & 0.77 \\
     U12082 & 10.89 $\pm$ 0.03 &  $--$  &  3.5 & 20.05 &  9.33 & 10.12 & 10.19 & 0.86 \\
     U12212 & 11.83 $\pm$ 0.04 &  $--$  &  2.8 & 20.02 & 11.27 & 10.78 & 11.39 & 0.25 \\
     U12632 & 10.41 $\pm$ 0.02 &  $--$  &  1.3 & 20.68 & 11.14 & 10.54 & 11.24 & 0.20 \\
     U12695 & 13.69 $\pm$ 0.10 &  $--$  & 10.0 & 20.82 &  8.95 &  9.36 &  9.50 & 0.72 \\
\enddata
\tablecomments{(1) Galaxy name. (2) apparent magnitude at 3.6$\mu$m. (3) $V-3.6$ color, corrected from $g$ for
SDSS data (4) Holmberg radius in kpc. (5) central 3.6$\mu$m surface brightness (mag arcsecs$^{-2}$). (6) stellar
mass.  (7) gas mass. (8) baryon mass (stars+gas). (9) gas fraction. }
\end{deluxetable}

\begin{deluxetable}{lcccccccc}
\tablecolumns{14}
\small
\tablewidth{0pt}
\tablecaption{HSB Photometric Properties}
\tablehead{
\colhead{Object} & \colhead{$m_{3.6}$} & \colhead{$V-3.6$} & \colhead{$R_{Holmberg}$} & \colhead{$\mu_o$} & 
\colhead{log $M_*$} & \colhead{log $M_{gas}$} & \colhead{log $M_{baryons}$} & \colhead{$f_g$} \\
\colhead{} & \colhead{} & \colhead{(kpc)} & \colhead{(kpc)} & \colhead{($M_{\sun}$)} & \colhead{($M_{\sun}$)} & \colhead{($M_{\sun}$)} & \colhead{} \\
\colhead{(1)} & \colhead{(2)} & \colhead{(3)} & \colhead{(4)} & \colhead{(5)} & \colhead{(6)} & \colhead{(7)} & \colhead{(8)} \\
}
\startdata
ESO563-G021 &  8.43 $\pm$ 0.01 &  $--$  & 43.1 & 15.24 & 11.27 & 10.78 & 11.39 & 0.25 \\
       N801 &  9.05 $\pm$ 0.01 &  $--$  & 46.2 & 16.65 & 11.13 & 10.52 & 11.22 & 0.20 \\
      N1003 &  8.92 $\pm$ 0.01 &  $--$  &  8.6 & 16.98 &  9.44 &  9.85 &  9.99 & 0.72 \\
      N1167 &  8.23 $\pm$ 0.01 &  $--$  & 40.7 & 17.07 & 11.33 & 10.13 & 11.36 & 0.06 \\
      N2998 &  9.39 $\pm$ 0.01 &  3.77$\pm$0.18  & 34.5 & 18.28 & 10.91 & 10.59 & 11.08 & 0.33 \\
      N5533 &  8.39 $\pm$ 0.01 &  3.98$\pm$0.13  & 38.1 & 17.46 & 11.19 & 10.52 & 11.27 & 0.18 \\
      N6195 &  9.80 $\pm$ 0.01 &  4.07$\pm$0.03  & 53.3 & 17.90 & 11.28 & 10.47 & 11.35 & 0.13 \\
      N6674 &  8.40 $\pm$ 0.01 &  $--$  & 12.0 & 16.75 & 11.06 & 10.73 & 11.23 & 0.32 \\
      U1551 & 10.21 $\pm$ 0.01 &  $--$  & 13.1 & 17.98 &  9.95 &  9.87 & 10.21 & 0.45 \\
      U2259 & 10.36 $\pm$ 0.02 &  $--$  &  2.1 & 18.47 &  8.30 &  8.33 &  8.62 & 0.52 \\
      U2885 &  8.48 $\pm$ 0.01 &  $--$  & 51.0 & 16.32 & 11.40 & 10.59 & 11.46 & 0.14 \\
      U2953 &  5.93 $\pm$ 0.01 &  $--$  & 16.5 & 15.92 & 10.72 &  9.71 & 10.76 & 0.09 \\
      U5709 & 12.42 $\pm$ 0.01 &  $--$  & 15.7 & 18.54 &  9.95 & 10.01 & 10.28 & 0.53 \\
     U11455 &  8.81 $\pm$ 0.01 &  $--$  & 33.1 & 14.53 & 11.20 & 10.39 & 11.27 & 0.13 \\

\enddata
\tablecomments{(1) Galaxy name. (2) apparent magnitude at 3.6$\mu$m. (3) $V-3.6$ color, corrected from $g$ for
SDSS data (4) Holmberg radius in kpc. (5) central 3.6$\mu$m surface brightness (mag arcsecs$^{-2}$). (6) stellar
mass.  (7) gas mass. (8) baryon mass (stars+gas). (9) gas fraction. }
\end{deluxetable}

\end{document}